\documentclass[aps,pra,twocolumn,showpacs]{revtex4-1}

\usepackage{graphicx}
\usepackage{color}
\usepackage{amsmath}
\usepackage{epstopdf}

\newcommand{\bom}{\boldsymbol{\omega}}

\begin{document}

\title{Diffusion of quantum vortices}

\author{E. Rickinson$^1$}
\author{N. G. Parker$^1$}
\author{A. W. Baggaley$^1$}
\author{C. F. Barenghi$^1$}

\email{carlo.barenghi@newcastle.ac.uk}

\affiliation{$^1$Joint Quantum Centre (JQC) Durham--Newcastle, 
School of Mathematics, Statistics and Physics, Newcastle University, 
Newcastle upon Tyne, NE1 7RU, United Kingdom}

\date{\today}

\begin{abstract}
We determine the evolution of a cluster of quantum vortices
initially placed at the centre of a larger vortex-free region.
We find that the cluster spreads out spatially. This spreading motion
consists of two effects: the rapid evaporation of vortex dipoles from the 
cluster and the slow expansion of the cluster itself. The latter is 
akin to a diffusion process controlled by the quantum of circulation.
Numerical simulations of the Gross-Pitaevskii equation show that this 
phenomenon is qualitatively unaffected by the presence of sound waves, vortex annihilations,
and boundaries, and it should be possible to create it in the laboratory.
\end{abstract}


\maketitle
    
\section{Introduction}

Atomic Bose-Einstein condensates (BECs) which are tightly confined in one
direction provide an ideal physical setting for  
two-dimensional (2D) vortex dynamics.  As long as the confining
third dimension of the condensate is comparable with
the width of a vortex core, 
longitudinal excitations of the vortices (Kelvin waves) are
suppressed, and the dynamics becomes effectively 2D. 
Under this condition, BECs become similar to 
2D flows ruled by the classical Euler equation (viscous-fee and irrotational
with the exception of isolated vortex singularities) as described in
traditional textbooks of fluid dynamics.  
The remarkable properties of BECs 
arise from the existence of a macroscopic complex order parameter
$\psi({\bf x},t)=\vert \psi({\bf x},t) \vert e^{i \phi({\bf x},t)}$ which,
at sufficiently low temperatures, obeys the Gross-Pitaevskii equation (GPE). 
The GPE can be reformulated \cite{Primer} into classical
hydrodynamic equations for a compressible
Euler fluid with velocity 
${\bf v}({\bf x},t)= (\hbar/m)\nabla \phi({\bf x},t)$ and 
number density $n({\bf x},t)=|\psi({\bf x},t) |^2$, where $m$ is the 
mass of one atom, $\hbar=h/(2 \pi)$ and $h$ is Planck's constant. Since
the curl of a gradient is zero, the vorticity $\bom=\nabla \times {\bf v}$ 
is zero everywhere with the exception of points 
where $\psi=0$ around which the circulation is an integer multiple of
the quantum of circulation
$\kappa=h/m$.  Multiply charged vortices are unstable, therefore
vortices in 2D BECs are limited to circulation $\pm \kappa$.

Atomic BECs are a fruitful field of investigations into vortex dynamics.
Landmark studies include the formation of 
Abrikosov lattices of same-signed vortices \cite{Hodby2001,AboShaeer2001}, 
states of quantum turbulence  \cite{Henn2009,Neely2013,Kwon2014,Reeves2012,Parker2005,Gasenzer2017},  
turbulence decay \cite{Kwon2014,Stagg2015,Cidrim2016,Groszek2016,Baggaley2018}, 
clustering \cite{White2012,Simula2014}, states of negative absolute
temperatures \cite{Kraichnan1980,Viecelli1995,Simula2014,Billam2014,Gauthier2018,Johnstone2018},
the Kibble-Zurek mechanism forming vortices  
as topological defects of the condensate phase 
\cite{Weiler2008,Lamporesi2013,Chomaz2015,Donadello2016,Beugnon2017}, vortex reconnections \cite{Serafini2017},
the creation of vortex dipoles
pairs \cite{Neely2010,Kwon2015} and
quantum analogues of classical phenomena, e.g. K\'{a}rm\'{a}n vortex 
streets \cite{Sasaki2010,Stagg2014,Kwon2016} and hydrodynamic instabilities \cite{BagPark2018,Kanai2018}.

Here we consider a distinct and fundamental behaviour of quantum 
vortices not considered to date:
how quantum vortices diffuse, that is, how they spread out from some initially localized region to fill the system. 
The analogy with classical flows is instructive.
In any classical 2D flow ${\bf v}(x,y,t)$
in the $xy$ plane, the vorticity $\bom=\omega_z \hat{\bf z}$ 
points in the $\pm z$ direction. If the fluid is also viscous, the classical
2D Navier-Stokes equation requires that $\omega_z$ obeys 
the vorticity equation 
\begin{equation}
\frac{{\rm D}\omega_z}{{\rm D}t} = \nu \nabla^2 \omega_z,
\label{eq:vort-diff}
\end{equation}
\noindent
where ${\rm D}/{\rm D}t=\partial/\partial t + ({\bf v} \cdot \nabla)$ is
the material derivative, and the kinematic viscosity $\nu$ plays the role
of diffusion coefficient. 
In a 2D BEC, $\omega_z$ consists of delta-functions of fixed strength.
Since $\nu=0$ for a superfluid, at first glance at 
Eq.~(\ref{eq:vort-diff}) suggests that quantum vorticity should not diffuse.
The aim of this work is to show that this not the case. By numerically
tracking the evolution of a large number of quantum vortices of equal 
positive and negative circulation, initially localized at random positions
within a small region, we shall find that the region of quantum vorticity
diffuses (spreads out) with time. We shall also determine the effective 
diffusion coefficient $\nu$, and find that it is of the order of magnitude
of the quantum of circulation $\kappa$. In other words, the classical
property of diffusion {\it emerges} from the mutual
interactions of quantum vortices. 
Finally, we shall demonstrate that we observe the same qualitative 
behaviour in the presence of boundaries, vortex annihilation and waves. 
Hence, it should be possible to demonstrate
the diffusion of quantized vorticity experimentally in an atomic condensate.

\section{Methods}
We use two methods: the point vortex method (PVM) and the GP equation (GPE).
The classical context of point vortices in a 2D incompressible inviscid 
fluid \cite{Saffman1992} is simpler than physical reality 
but, as already remarked \cite{Middelkamp2011,Simula2014,Billam2014},
it accounts for the essential ingredient:  vortex interactions.

The GPE provides an accurate description of 
superfluid dynamics at temperatures much less than the critical 
temperature \cite{Freilich2010,Allen2013}. Besides vortex interactions,
the GPE accounts for effects such as
finite-sized vortex cores, vortex annihilations when vortices collide,  
sound emission when vortices accelerate, and 
the presence of experimentally realistic boundaries, which are not present in
the point vortex method. 

Using either the PVM or the GPE, the initial condition of all our numerical
simulations consists of a cluster of $N$ vortices 
(half of positive, half of negative circulation) 
placed at random locations near the centre of the system sampled from a 
2D multivariate normal distribution with standard 
deviations $\sigma_x=\sigma_y=\sigma_0$. Typical values used range from
$N=100$ (for GPE simulations and vortex point simulations
inside circular domains) to $N=500$ (for vortex point simulations 
in the infinite plane). Vortices in the cluster are observed to spread through two processes (See Sec. III (A) for details): the gradual spatial diffusion of the cluster and the evaporation of vortex dipoles from the cluster spread; only the vortices spreading through the first process are relevant to our analysis, and their number is denoted by $N_0$. Note that our values of $N$ 
are much larger that the number of 
vortices required for chaotic dynamics ($N=4$ in an infinite plane 
and $N=3$ in a circular domain \cite{Aref2007}).
The width of the initial vortex cluster is taken to be much less than 
the size of the system (we typically take $\sigma_0=10$ in the 
dimensionless units we employ - see later).

In all simulations we let the vortex
configuration to evolve in time and monitor
the root mean square deviation of vortices from their initial positions, 
defined as
\begin{equation}
d_{\text{rms}}(t)=\sqrt{ \frac{1}{N_0(t)}\sum^{N_0(t)}_{i=1}\left( 
\Delta x_i^2(t)+\Delta y_i^2(t) \right)}
\label{eq:drms}
\end{equation}

\noindent
where $\Delta x_i(t)=x_i(t)-x_i(0)$ and $\Delta y_i(t)=y_i(t)-y_i(0)$.

\subsection{The point vortex method}

The equations of motion of $N$ point vortices of circulation 
$\kappa_j$ and position $x_j(t),y_j(t)$ are
\begin{eqnarray}
\frac{dx_j}{dt}=-\frac{1}{2 \pi} \sum_{k \neq j} \frac{\kappa_k (y_j-y_k)}{r_{j,k}^2},\nonumber \\
\frac{dy_j}{dt}=\frac{1}{2 \pi} \sum_{k \neq j} \frac{\kappa_k (x_j-x_k)}{r_{j,k}^2},
\label{eq:pvm}
\end{eqnarray}

\noindent
($j=1, \cdots N$) where $r_{j,k}=\sqrt{(x_j-x_k)^2+(y_j-y_k)^2}$. 
For point vortices \cite{Newton2001}, configuration space is the same 
as phase space and Eqs.~(\ref{eq:pvm}) are equivalent to Hamilton's 
equations $ \kappa_j dx_j/dt=\partial H/\partial y_j$ and 
$\kappa_j dy_j/dt=-\partial H/\partial x_j$, where the energy is the
Hamiltonian
\begin{equation}
H=-\frac{1}{4 \pi} \sum_j \sum_{k \neq j} \kappa_j \kappa_k \ln{(r_{j,k})}.
\label{eq:H}
\end{equation}

We solve Eqs.~(\ref{eq:pvm}) numerically using a $6^{\rm th}$-order 
Runge-Kutta scheme with typical time-step $\Delta t=10^{-3}$. In
a typical simulation of $N=500$ vortices for $10^6$ time steps, energy
is conserved with relative percent error
$100 \times \vert H(t)-H(0)\vert/H(0) \approx 3 \%$.

To simulate point vortices within a disc of radius $a$ 
we use the method of images. Since we have an equal number of
positive and negative vortices, 
each vortex at position $(x_j,y_j)$ has an image vortex 
(of opposite circulation) at $(x'_j,y'_j)$ where \cite{Newton2001}
\begin{equation}
x'_j=\frac{a^2 x_j}{(x^2_j+y_j^2)},
\qquad
y'_j=\frac{a^2 y_j}{(x^2_j+y_j^2)},
\end{equation}

\noindent
resulting in the following equations of motion
\begin{eqnarray}
\frac{dx_j}{dt}=-\frac{1}{2 \pi} \sum_{k \neq j} \frac{\kappa_k (y_j-y_k)}{r_{j,k}^2}
+\frac{1}{2 \pi} \sum_k \frac{\kappa_k (y_j-y'_k)}{r_{j,k}^2},\nonumber \\
\frac{dy_j}{dt}=\frac{1}{2 \pi} \sum_{k \neq j} \frac{\kappa_k (x_j-x_k)}{r_{j,k}^2}
-\frac{1}{2 \pi} \sum_k \frac{\kappa_k (x_j-x'_k)}{r_{j,k}^2};
\label{eq:pvm-disc}
\end{eqnarray}

\noindent
The images guarantee that the flow has zero tangential velocity component 
at the boundary.

\subsection{The GPE method}

We write the GPE in dimensionless form using $\xi=\hbar/\sqrt{m \mu}$ 
(the healing length) as unit of distance, $\tau=\hbar/\mu$ as unit of time, 
and $n_0=\mu/g$ as units of density,  where $\mu$ is the chemical potential,
$g$ the (2D) interaction parameter \cite{Primer}, and $m$ the mass of one atom. 
We obtain the following 
(dimensionless) equation for the (dimensionless) wavefunction $\psi(x,y,t)$
\begin{equation}
i  \frac{\partial \psi }{\partial t}=-\frac{1}{2}\nabla^2 \psi + \psi \vert \psi \vert^2 - \psi + V \psi
\label{eq:GPE}
\end{equation}
\noindent
which we solve in the (dimensionless) numerical domain $-L \le x,y, \le L$ with
zero boundary conditions at the edge (typically $L=204.8$). 
The (dimensionless) confining potential $V(x,y)$ is a box trap with 
sharp potential walls; such box traps can be produced experimentally 
using optical or electromagnetic 
fields \cite{Meyrath2005,vanEs2010,Gaunt2013,Chomaz2015}.  
Within the box, where the potential is negligible, the 
condensate's density is constant, as in the point vortex model.  

We focus on a circular box trap with potential defined as
\begin{equation}
V(x,y) = \frac{1}{2}V_0\left( \frac{r}{R_0} \right)^{\alpha},
\label{eq:box}
\end{equation}

\noindent
where $r=\sqrt{x^2+y^2}$, $R_0$ is the box trap's effective (dimensionless)
radius, and the exponent $\alpha$ determines the steepness of the trap's wall.
Typically we take $R_0=200$ and $\alpha$ ranges from $15$ to $100$.
Equation~(\ref{eq:GPE}) is solved using a $2^{\rm{nd}}$-order finite 
difference method and a $4^{\rm{th}}$-order Runge-Kutta time stepping 
scheme; typical grid size and time step are $\Delta x=\Delta y=0.2$ 
and $\Delta t=10^{-2}$. To find the ground state we start from
the Thomas-Fermi profile \cite{Primer}
\begin{equation}
\displaystyle
\psi(x,y)=\begin{cases}
    \sqrt{1-V(x,y)}, & \text{if $V(x,y)<1$}.\\
    0, & \text{otherwise},
  \end{cases}
\end{equation}

\noindent
which is evolved in imaginary time \cite{Bader2013} typically 
for $10^4$ time steps, renormalizing $\psi$ at each step to preserve
the number of atoms, yielding the ground state solution. 
To imprint a vortex in the ground state
at position $(x_0,y_0)$, we impose a $2\pi$ phase 
winding around $(x_0,y_0)$, and use a $2^{\rm{nd}}$-order Pad\'e 
approximation \cite{Caliari2016} as an initial density profile. 
Equation~(\ref{eq:GPE}) is then propagated for a further $100$ time steps 
in imaginary time, again renormalizing $\psi$ at each time step, 
in order to find the correct density profiles of the vortex.

\section{Results}

\subsection{Evaporating dipoles and a diffusing cluster}
\label{sec:3a}

We first analyse the results of numerical simulations of the point vortex 
model in free space, Eq.~(\ref{eq:pvm}).

During the time evolution, we observe that 
the initial vortex cluster spreads in space via two different processes. 
The first process is a gradual spread of the cluster which arises 
from the mutual interaction of vortices. The second process is 
the evaporation of vortex dipoles from the cluster; 
a vortex dipole is a pair comprising a vortex ($\kappa_j=1$) 
and an antivortex ($\kappa_j=-1$) which, in isolation, would propagate 
at constant speed along a straight line. The evaporation effect,
reported to explain superfluid helium experiments \cite{BS2002}, is easily
understood.
Let $\delta$ be the distance between the vortex and the antivortex 
of a pair, and $\ell \approx n^{-1/2}$ be the average distance 
between vortices in the cluster where $n \approx N/(\pi \sigma_0^2)$ 
is the initial number density of vortices in the cluster. If a 
pair with $\delta< \ell$ is sufficiently close to the edge of the 
cluster and is directed outward, it will escape from the cluster. 
The probability that the pair is re-captured by the cluster is small, 
because its translational speed, $\kappa/(2 \pi \delta)$, is larger than 
the average vortex velocity within the cluster, $\kappa/(2 \pi \ell)$. 
In general, the energy of a vortex dipole increases 
with $\delta$. Therefore, the formation and evaporation of vortex dipoles 
allows the vortex cluster to expand in space while keeping 
the total energy constant.

\begin{figure}[h]
\includegraphics[width=0.7\columnwidth]{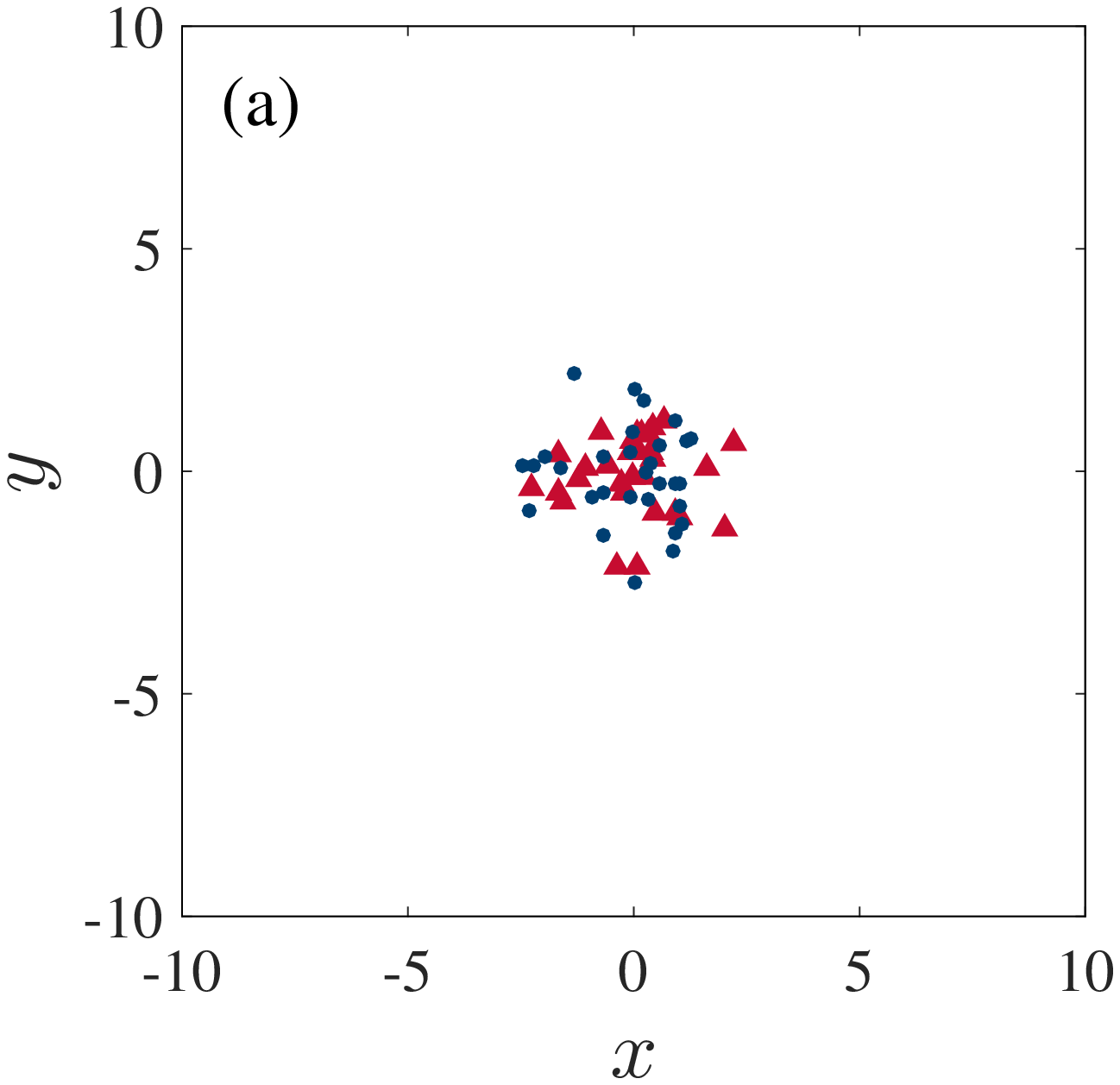} 
\includegraphics[width=0.7\columnwidth]{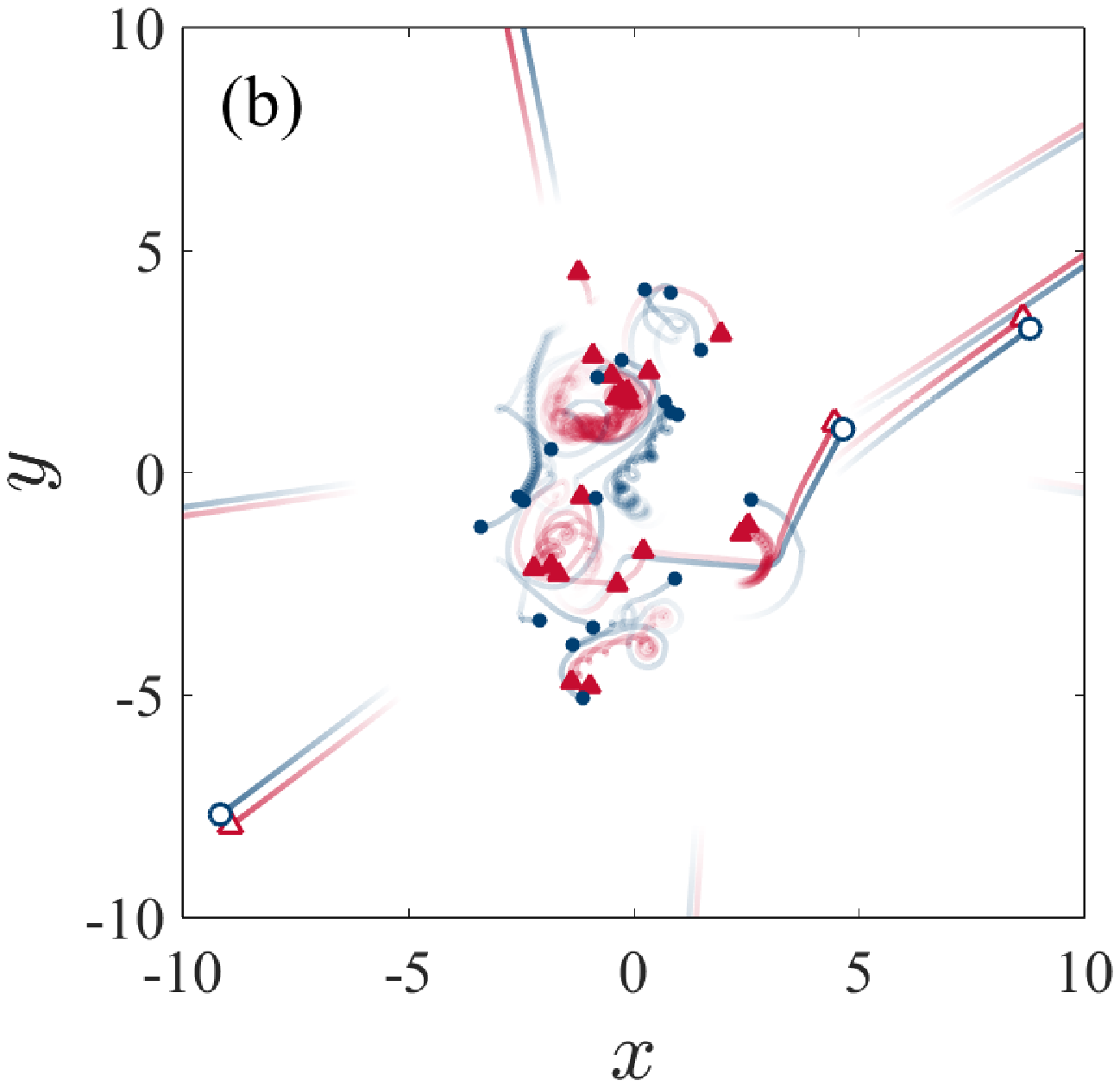} 
\caption{Point-vortex simulation (with a low number of vortices
for clarity). Snapshots of vortex configurations 
at $t=0$ (a) and $t=20$ (b). 
The range of the display ($-10 \le x,y \le 10$) is for visualization only
(the computational domain is the infinite plane).  
Vortices (anticlockwise rotation) and antivortices (clockwise circulation) 
are marked with solid red triangles and solid blue circles respectively. 
Vortices comprising dipoles (vortex-antivortex pairs)
are marked with hollow triangles and circles.
The fading comet tails visualize vortex trajectories thus helping the 
identification of vortex dipoles which `evaporate' from the cluster.
}
\label{fig1}
\end{figure}

The two processes are qualitatively 
illustrated in Fig.~\ref{fig1} (with reduced number of 
vortices for clarity). The evaporating dipoles,  
marked in Fig.~\ref{fig1}(b) with hollow symbols, move away from the
cluster along straight trajectories (highlighted by the comet tails),
whereas vortices in the cluster move
along zig-zagging trajectories. The effect is described by the first
movie in the Supplementary Material \cite{supple1}. 

\begin{figure}[h]
\includegraphics[width=1\columnwidth]{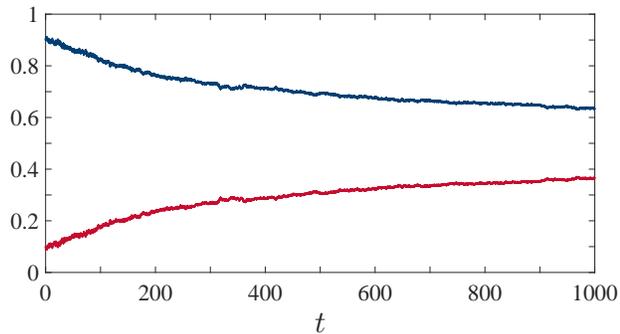}
\caption{Point-vortex simulation.
Fraction $N_0(t)/N$ of vortices that remain in the cluster 
(top blue curve), and fraction $N_1(t)/N$ of vortices that form dipoles 
(bottom red curve), plotted vs. time $t$ and averaged over 10 realizations.
}
\label{fig2}
\end{figure}

To distinguish the evaporation processes from the
spatial diffusion of the main cluster it is useful to divide the vortices
in two groups: the vortex dipoles and the main cluster.
This distinction requires a consistent definition of what a vortex dipole is.
We deem two vortices of opposite circulation to form a dipole if they satisfy the following criteria: (i) the separation $\delta$ between the vortices is less than the separation between either vortex and any other vortex; they are mutually closest to each other, (ii) $\delta$ is less than half the separation between either vortex and the nearest vortex of the same circulation. By numerically experimenting, we find that relaxing or restricting the threshold value for the second criteria by a factor of two has negligible effect on the identification of dipoles.
 At each time during the evolution we count the numbers $N_0(t)$ 
of vortices which are part of the cluster and the number $N_1(t)$ 
of vortices which are evaporating dipoles. Since in the 
point vortex model vortices cannot annihilate, the total number of 
vortices, $N=N_0(t)+N_1(t)$, remains constant. 
Figure~\ref{fig2} shows the relative 
fractions $N_0/N$ and $N_1/N$ as a function of time. Notice
that approximately $20\%$ of the vortices are identified as vortex dipoles 
already at $t=0$, and that the rate at which vortex dipoles form decreases 
with time, with approximately 1/3 of vortices having become
dipoles when the simulation is stopped.

\begin{figure}[h]
\includegraphics[width=1\columnwidth]{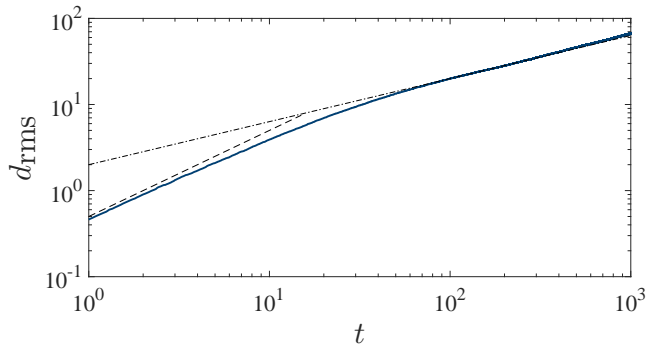}
\caption{Point-vortex simulation. Root mean square deviation of vortices 
from their initial position, $d_{\text{rms}}$, vs. time, $t$. 
Only the vortices which are identified as part of the cluster are used 
in the calculation of $d_{\text{rms}}$. 
The dashed and dot-dashed lines correspond to $t$ and 
$t^{1/2}$ scalings respectively.
}
\label{fig3}
\end{figure}

By restricting our attention to the $N_0(t)$ vortices which remain 
in the cluster, we calculate the root mean square deviation of vortices
from their initial positions, $d_{rms}$, averaging over 10 realizations.
Figure~\ref{fig3} shows that the initial linear spread,
$d_{\text{rms}} \sim t$,  is followed by $d_{\text{rms}} \sim t^{1/2}$ 
at large $t$, the typical scaling of diffusion processes \cite{Uhlenbeck1930}. 
Similarly to Brownian motion, in which particles follow linear trajectories 
before colliding with other particles, initially vortices 
move away at constant speed from their starting positions, 
leading to the  linear (ballistic) spread, until the direction of their 
motion is altered by approaching another vortex close enough for its
contribution to the velocity field to become dominant; a succession of such 
events appears to behave as a random walk. This interpretation is supported by the observation that the change in behaviour from $d_{\text{rms}} \sim t$ to $d_{\text{rms}} \sim t^{1/2}$ occurs when $d_{\text{rms}}\sim\sigma_0$.

\subsection{Point vortices in a disc}
\label{sec:3b}

To make the point vortex model more realistic and include the
presence of the condensate's boundary, we perform numerical 
simulations within a circular disc of radius $a$ 
using Eq.~(\ref{eq:pvm-disc}), rather than Eq.~(\ref{eq:pvm}).
The scenario which we have described 
(evaporating dipoles and slowly diffusing cluster)
does not change, but a new feature arises: 
when a vortex dipole approaches the disc's boundary, the 
distance $\delta$ between the vortex and the antivortex
increases to conserve energy, until, close to the boundary,
the vortex pair breaks up. The vortex and the antivortex, now separated,
travel in opposite directions along the circular boundary, driven by their 
images. Consider for example the fate of the vortex: 
as it travels along the boundary,
eventually it will meet an antivortex (travelling along the 
boundary in opposite direction) resulting from the break-up of a different
vortex dipole, which has reached the boundary too; when the distance $\delta$
between vortex and antivortex becomes less than the distance
to the boundary, the vortex and the antivortex make a sharp turn together,
forming a new vortex dipole which travels radially towards  
the centre of the disc. A similar fate is met by the antivortex.
In other words, the boundary effectively re-injects evaporating
vortex dipoles back against the main cluster. This effect complicates the
interpretation of the results. Clearly the effect is absent if
the radius of the disc, $a$, is much larger than the size of the
initial vortex cluster, $\sigma_0$, and we limit our analysis to
times shorter than the time taken by evaporating vortex dipoles 
to reach the boundary. However there are experimental limitations
on the size of a 2D condensate, and the question remains as to whether
the reflection of vortex dipoles by the boundary revealed by the
point vortex model is physically realistic, and whether it prevents 
clear separation between evaporating dipoles and main cluster in
an actual experiment; we shall have to return to this
effect when discussing the GPE model.

\subsection{The GPE model}

We now turn the attention to the GPE model in which the condensate 
is confined within a box trap. Here we present only results obtained
using circular box traps, but we have repeated the simulations in 
square box traps and find the same qualitative behaviour, indicating that 
our results are robust to the trap geometry.  

\begin{figure*}[h]
\centering
\includegraphics[width=0.4\linewidth]{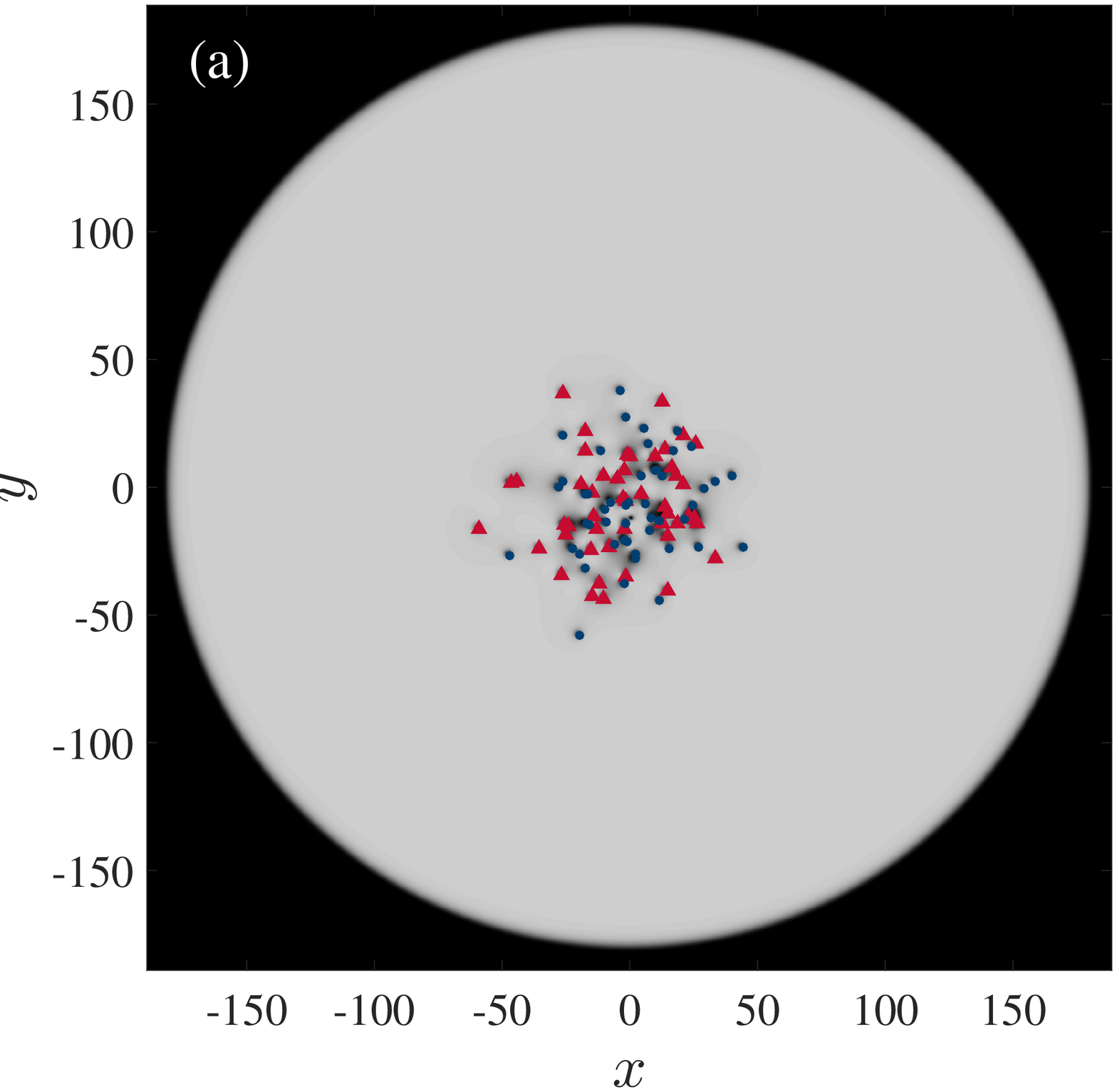}\hspace{5pt}\includegraphics[width=0.4\linewidth]{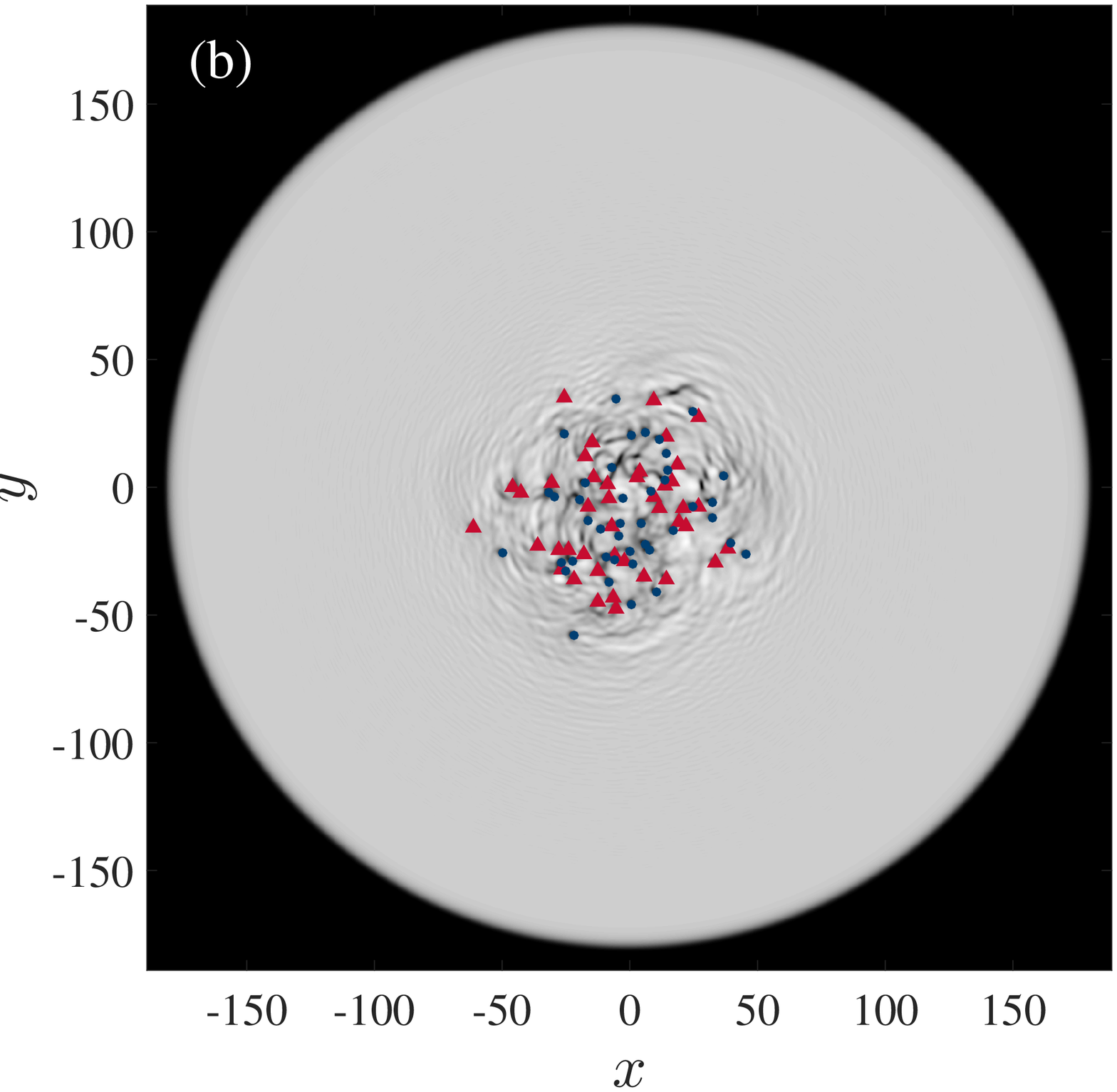}\\
\includegraphics[width=0.4\linewidth]{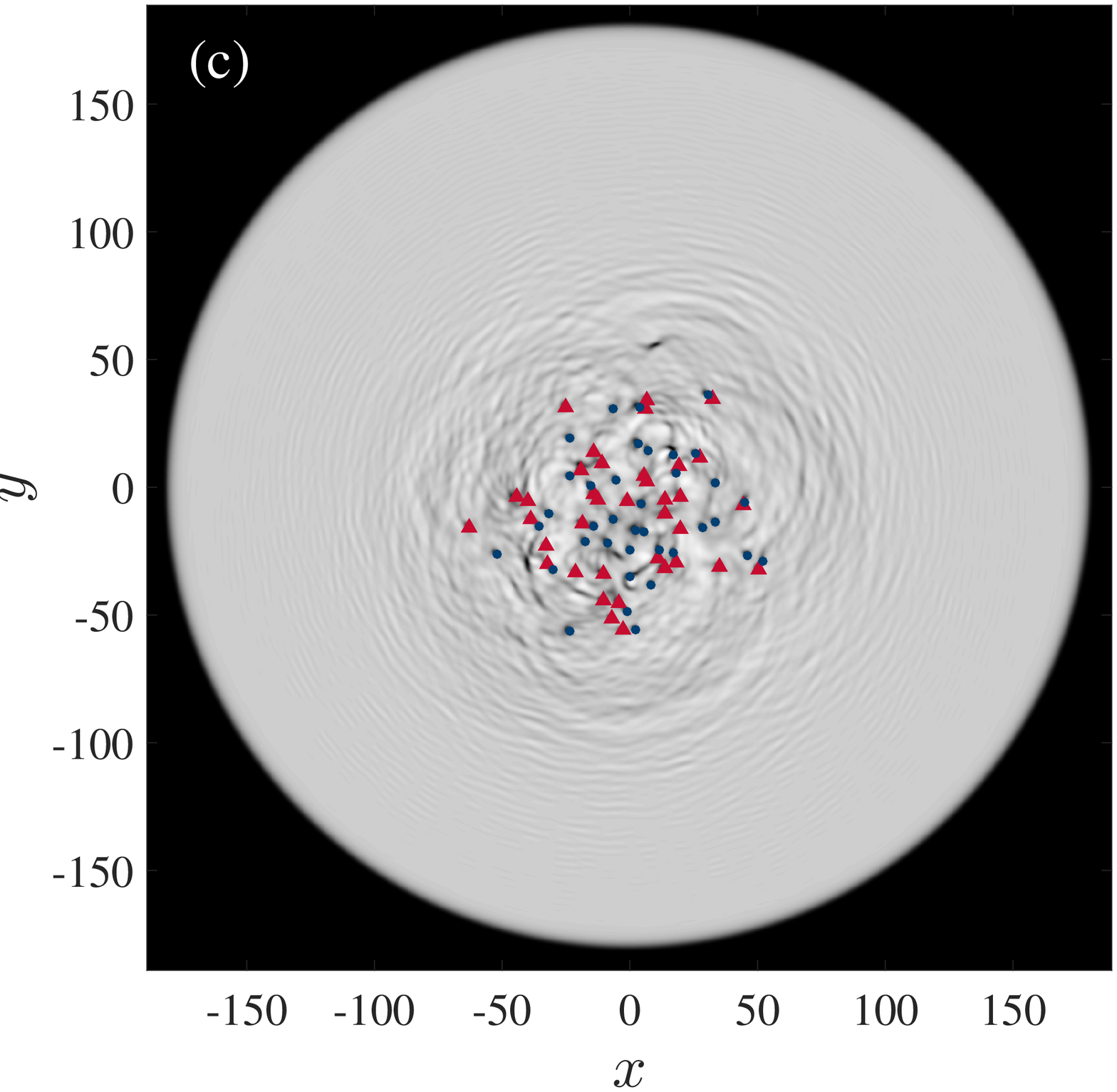}\hspace{5pt}\includegraphics[width=0.4\linewidth]{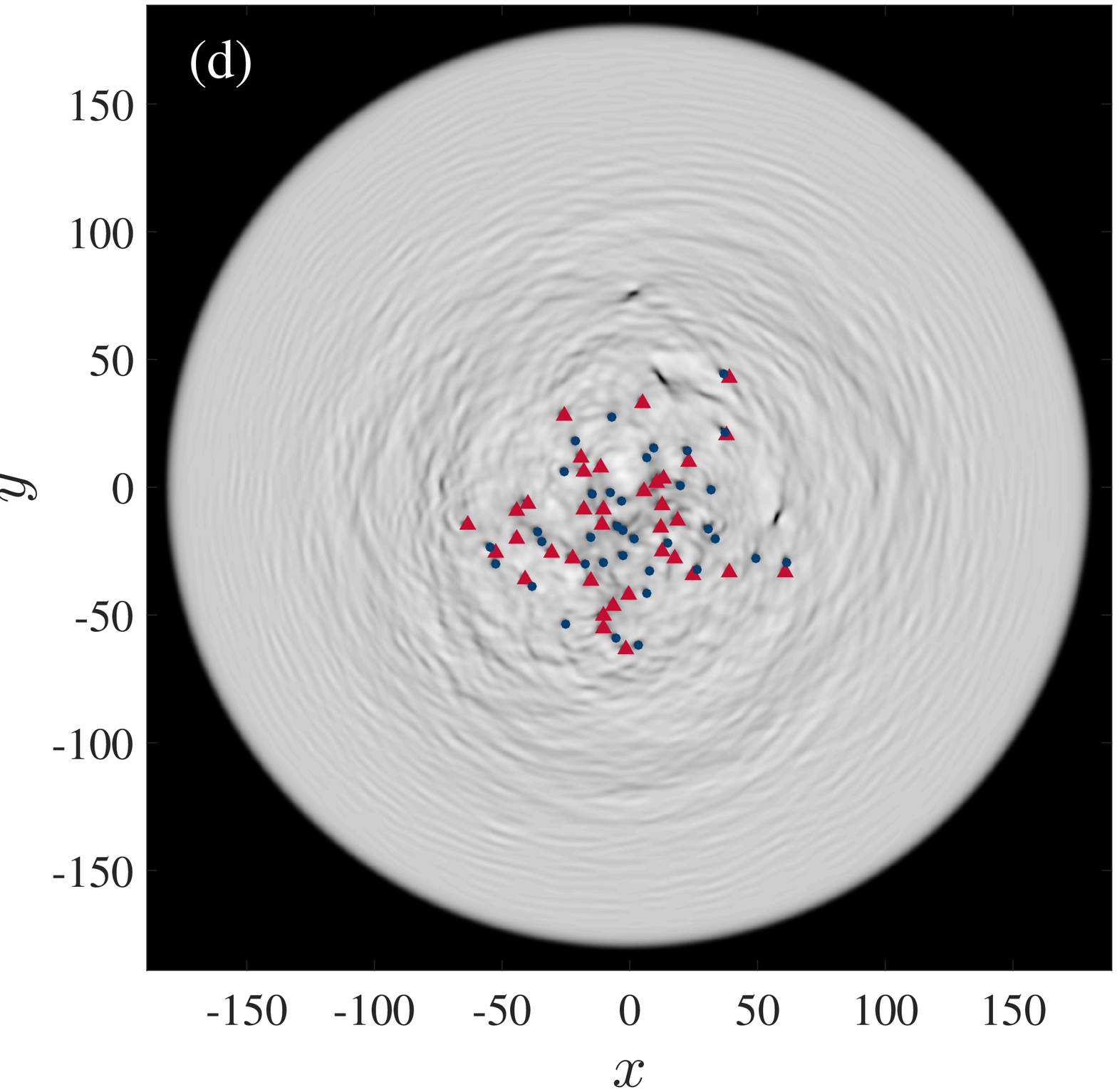}\\
\includegraphics[width=0.4\linewidth]{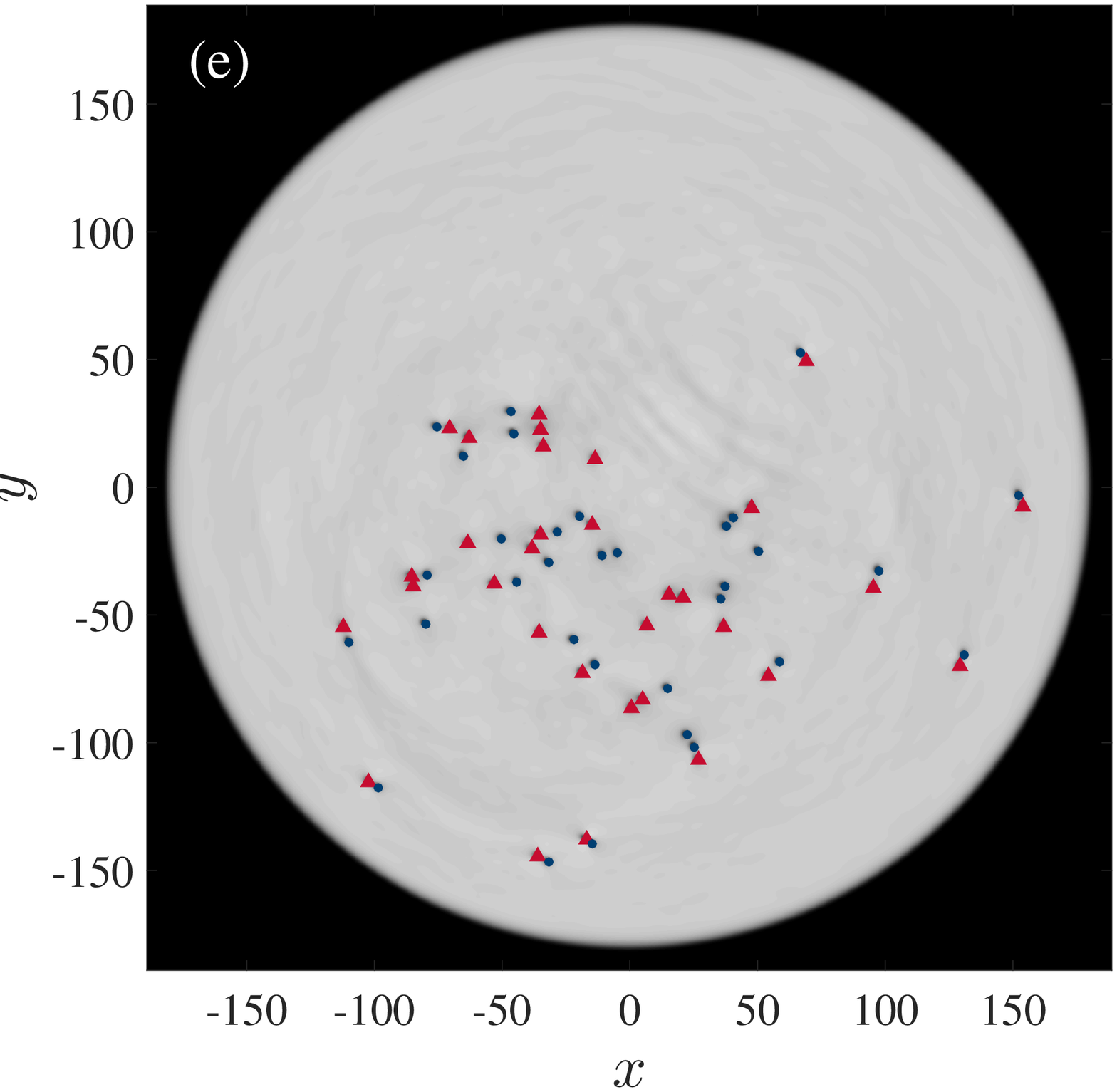}\hspace{5pt}\includegraphics[width=0.4\linewidth]{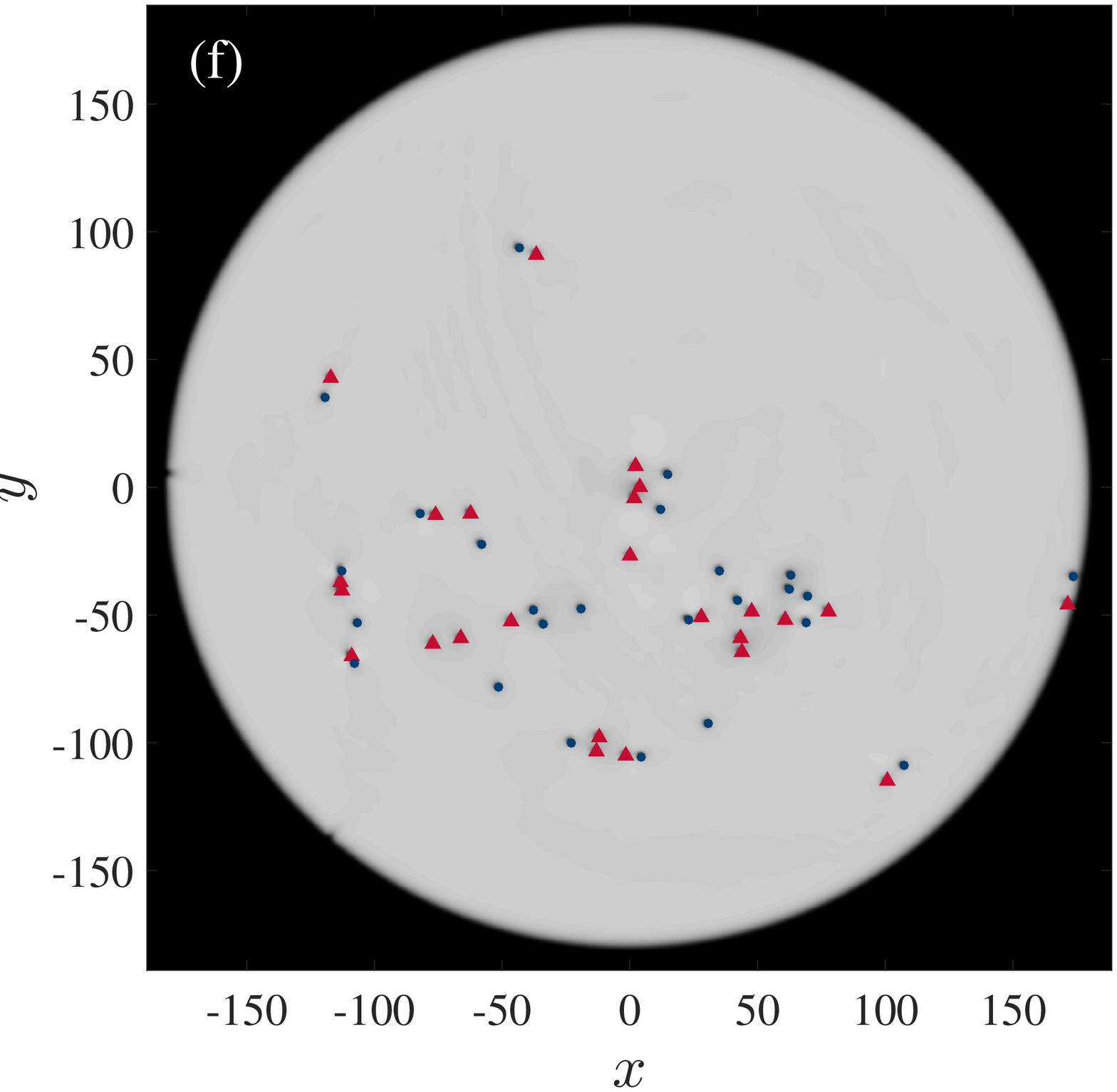}\\
\hspace{28pt}\includegraphics[width=0.78\linewidth]{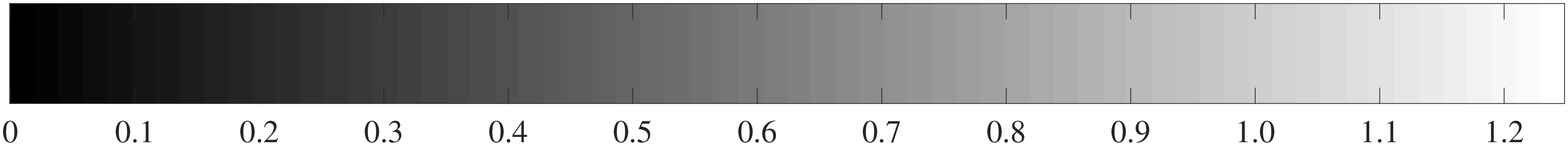}
\caption{GPE simulation (circular box trap). 
Plots of the condensate's density $\vert \psi \vert^2$ on the $x,y$ plane 
at $t=0$ (a), $t=30$ (b), $t=60$ (c), $t=90$ (d), $t=500$ (e) and $t=1000$ (f). 
Positive and negative vortices are marked by solid red triangles and 
solid blue circles respectively. Note the sound waves generated by vortices,
particularly in panels (b), (c) and (d).}
\label{fig4}
\end{figure*}

\begin{figure}[h]
\includegraphics[width=1\columnwidth]{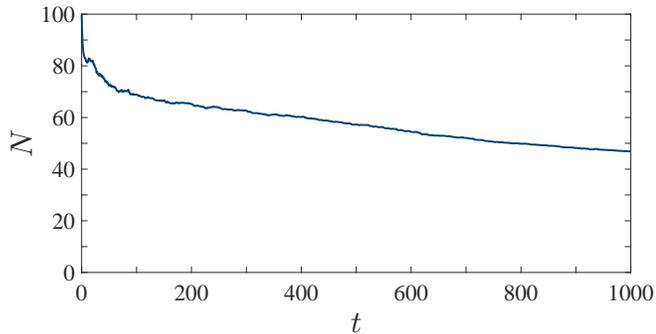}
\caption{GPE simulations (circular box trap). Number of vortices $N$ at time $t$ averaged over $20$ simulations.
} 
\label{fig5}
\end{figure}

The typical evolution of the initial vortex configuration
is shown in Fig.~\ref{fig4}, where we plot
the density of the condensate at successive selected times. 
The dark spots in the figure are the  density-depleted vortex cores, 
which we label with red triangles and blue circles (as in Fig.~\ref{fig1}
to distinguish positive from negative vortices). 
As for the point vortices, we observe that fast vortex dipoles evaporate from the slower spreading vortex cluster.
Unlike point vortices, however, vortices in the GPE model excite 
density waves when they accelerate \cite{Leadbeater2003,Parker2005}.
These waves are the ripples clearly visible in Fig.~\ref{fig4},
particularly at early stages when vortices are closer to each other
and undergo more rapid velocity changes. Moreover, unlike the point vortex 
model, the total number of vortices is not constant, 
as shown in Fig.~\ref{fig5}, because some vortices collide and
annihilate during the real time evolution, particularly again at early stages, when vortices are closer to each other.

\begin{figure}[!h]
\includegraphics[width=1\columnwidth]{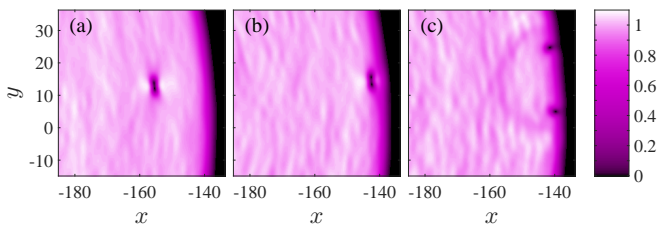}
\caption{GPE simulation in a circular box trap. 
Density snapshots are at (a) $t/\tau = 250$, (b) $t/\tau = 275$, 
(c) $t/\tau=300$, with $x$ and $y$ in units of $\xi$. A vortex dipole 
is seen approaching the boundary in (a), meeting the boundary in (b), 
separating into constituent vortex and antivortex which travel 
along the boundary in (c), emitting sound waves in this interaction. 
The colour scale is not linear but chosen to highlight small 
density perturbations. A movie \cite{supple2} shows this interaction in full.}
\label{fig6}
\end{figure}

Figure~\ref{fig6} shows that vortices which reach the boundary
of the condensate are re-injected back, exactly as found by the point vortex
model in Section~\ref{sec:3b}. 
Figure~\ref{fig6} illustrates the effect in a box trap with $\alpha=100$ and 
$R_0=200$; similar results are found with less steep confinement, 
e.g. $\alpha=15$.  We think that this re-injection takes place only at the
lowest temperatures described by the GPE.
Finite-temperature effects cause friction between the vortices and 
the cloud of thermal atoms. For example, an isolated vortex spirals
out of the condensate \cite{Jackson2009} and a vortex dipole
shrinks and vanishes, generating a Jones-Roberts soliton.
We know that thermal atoms are concentrated near the trap's boundary 
where the condensate's density vanishes \cite{Proukakis2008,Jackson2009}.
Therefore, even at relatively low temperature, we expect that vortices 
will suffer friction in the vicinity of the boundary.
This friction
can be modelled by replacing $i \partial/\partial t$ at
the left-hand side of Eq.~(\ref{eq:GPE}) 
with $(i-\gamma) \partial /\partial t$ 
where $\gamma$ is a small phenomenological parameter which we make 
radius-dependent by setting

\begin{figure}[!h]
\begin{center}
\includegraphics[width=1\columnwidth]{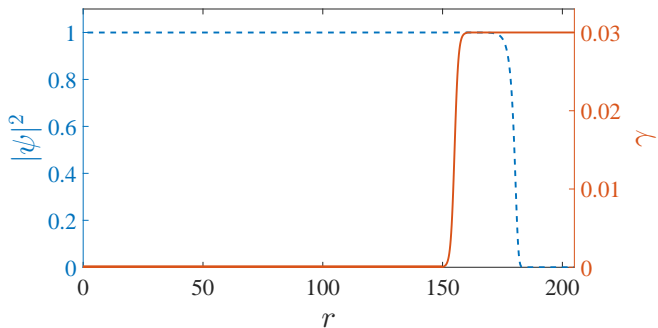}
\end{center}
\caption{GPE simulation in a circular box trap with localized damping. Density $\vert\psi\vert^2$ (dashed blue line) and phenomenological damping $\gamma$ (solid red line) are plotted against radius. Note the separate $y$-axis scales for each line.}
\label{fig7}
\end{figure}

\begin{figure}[!h]
\begin{center}
\includegraphics[width=1\columnwidth]{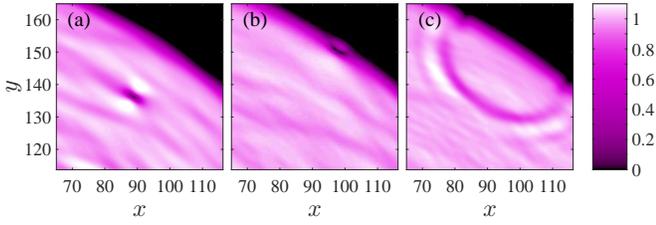}
\end{center}
\caption{GPE simulation in a circular box trap with localized 
damping near boundaries. Density snapshots are at (a) $t/\tau = 225$, 
(b) $t/\tau = 250$, (c) $t/\tau=275$, with $x$ and $y$ in units of $\xi$. 
A dipole is seen approaching the boundary in (a), annihilates with the 
boundary in (b), producing a sound wave which can be seen spreading 
back into the condensate in (c). A movie \cite{supple3} shows 
this interaction in full.}
\label{fig8}
\end{figure}

\begin{center}
\begin{math}
\displaystyle
\gamma(r)=\begin{cases}
    0, & \text{if $0\leq r<150$}.\\
   \dfrac{\gamma_c}{2}\left[1+\tanh\left\{\pi(r'-1)\right\}\right], & \text{if $150\leq r <160$},\\
    \gamma_c,& \text{otherwise},
  \end{cases}
\end{math}
\end{center}

\noindent
where $r'=(r-150)/\beta$, $\beta = 5$ controls the width of the transition from $\gamma = 0$ to $\gamma = \gamma_c$, and $\gamma_c=0.03$ (the value inferred from experiments and commonly used
\cite{Choi1998,Tsubota2001}), with the tanh function providing a 
smooth transition between regimes (a step function may reflect vortices 
and sound waves). The profile of the localized damping is shown with the density profile of the condensate in Fig.~\ref{fig7}. With this finite-temperature correction, 
we find that most vortex dipoles annihilate at the boundary, producing sound 
waves that spread back into the bulk of the condensate without affecting
much the spread of the vortex cluster, as shown in Fig.~\ref{fig8}.

\subsection{Effective diffusion}

Using the point vortex model or the GPE model, 
we proceed with the analysis as initially described for point vortices. 
First we separate the fast-moving vortex dipoles from the vortices
which remain in the cluster behind which we further analyse.
By performing a number of simulations (typically 20 to 40),
we obtain the ensemble-averaged
root mean square deviation $d_{\text{rms}}(t)$ 
of the vortices in the cluster from their initial positions
as a function of time using Eq.~(\ref{eq:drms}). 
In the case of a scalar field
$f(x,y,t)$ which satisfies the 2D diffusion equation
\begin{equation}
\frac{\partial f}{\partial t}=\nu\, \nabla^2f,
\end{equation}

\noindent
with constant diffusion coefficient $\nu$, the
root mean square deviation $d_{\text{rms}}(t)$ and $\nu$ are related by
\begin{equation}
\nu=\frac{d^2_{\text{rms}}(t)}{4t},
\label{eq:nuprime}
\end{equation}

\begin{figure}[h]
\centering
\includegraphics[width=1\columnwidth]{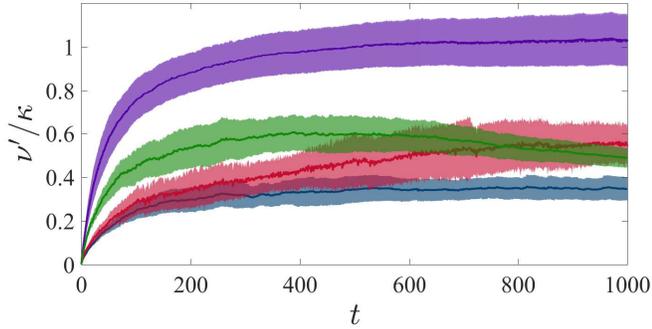}
\caption{
Ratio $\nu'/\kappa$ of effective diffusion coefficient 
and quantum of circulation for (top to bottom)
point vortex model in infinite domain (purple curve),
point vortex model in circular disc (green curve), GPE simulations in the circular box trap with damping near boundaries (blue curve), and in a square
box trap (red curve). The $95\%$ confidence intervals are indicated by the shaded regions.
}
\label{fig9}
\end{figure}

\begin{figure}[h]
\centering
\includegraphics[width=1\columnwidth]{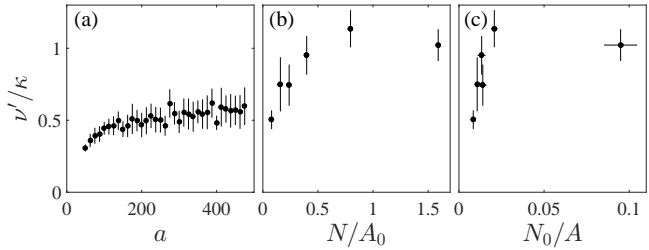}
\caption{
Ratio $\nu'/\kappa$ of effective diffusion coefficient and quantum 
of circulation for (a): point vortex model in a disc, vs. disc radius $a$ for $N/A_0=0.08$, (b): point vortex model in infinite domain, vs. initial number density $N/A_0$, where $A_0=\pi\sigma_0^2$ where $\sigma_0$ is the width of the initial vortex distribution, (c): point vortex model in infinite domain, vs. instanteneous number density $N_0/A$, where $N_0$ is the number of vortices remaining in the cluster and $A=\pi\sigma_x\sigma_y$, computed at the point at which $\nu^{\prime}/\kappa$ is computed. In all panels the data represent mean values with $95\%$ confidence intervals indicated by vertical (and in (c) horizontal) black lines.
}
\label{fig10}
\end{figure}

\noindent
Using Eq.~(\ref{eq:nuprime}), we define an effective diffusion
coefficient $\nu'$ representing the spatial spreading of the
vortex configuration. In order to combine results obtained
via the point vortex model and the GPE model, it is convenient 
to express $\nu'$ in terms of the quantum of circulation $\kappa$.
The results are presented in Fig.~\ref{fig9}
where we plot $\nu'/\kappa$ vs. $t$. We observe that, for large $t$,
$\nu'/\kappa$ tends to settle down to a constant.
For the point vortex model in the infinite domain we find that
$\nu'/\kappa \approx 1$, a result which is 
expected from dimensional arguments \cite{Tsubota2003}, but still 
surprising because the fluids which we have considered in our
models are inviscid. However, within the circular domain, 
Fig.~\ref{fig9} shows that $\nu' /\kappa \approx 0.5$ with both the
point vortex and GPE models, that is to say irrespective of 
the presence or the absence of density waves and vortex annihilations. 
We see similar behaviour in the GPE simulations if the domain is square, 
indicating that the phenomenon is not sensitive to the shape of
the confining geometry.
The reduced $\nu'/\kappa$ appears to be due to confinement itself (that is,
the image vortices, which are implicit in the GPE \cite{MasonBerloff2008}):
Fig.~\ref{fig10}, obtained using the point vortex model,
shows that $\nu'/\kappa$ increases as the radius $a$ of the disc increases, 
and that it tends to unity only if the initial vortex density $N/A$ is large
enough to provide sufficient scattering events of vortices.

With either the point vortex or the GPE
model, each initial condition (a set of random vortices of zero
net polarity placed in a prescribed initial region) has different
energy $E$ and angular momentum $L_z$ which are preserved by the
time evolution (with the exception of the GPE model with damping).
If we plot $\nu'/\kappa$ vs. $E$ or $L_z$ from all simulations
we observe a cloud of points with no significant dependence of $\nu'/\kappa$
vs. $E$ or $L_z$. 
Thus we conclude that our result do not depend on the initial condition,
although the spread in values of $E$ and $L_z$ may contribute to the
error bars of Fig.~\ref{fig9}.
An experiment to study the diffusion of vortices in a trapped atomic
condensate is also likely to start from vortices which are randomly
generated, as for our numerical simulations.

\section{Discussion}

In conclusion, both the point vortex model and the GPE model predict
that quantum vortices, if initially localized in a region,
spread out in space as a combination of two processes: the formation 
and fast evaporation of vortex-antivortex pairs (vortex dipoles), and 
the slower spread of the remaining vortex cluster. The latter, 
which can be modelled as a diffusion process, is insensitive to the
presence of density waves and annihilations, and depends essentially
on the Eulerian interaction between vortices which is captured by
a model as simple as the point vortex model. This is remarkable:
the classical diffusion of vorticity {\it emerges} in a fluid without
viscosity (a superfluid) from the interaction of many quantum vortices.
Quantitatively, the value of the effective 
diffusion coefficient $\nu'$ is in the range $0.3 <\nu'/\kappa < 1$,
essentially of the order of the quantum of circulation
as predicted by dimensional argument. 
In confined systems, the diffusion coefficient is reduced by the effect
of image vortices.

Both the evaporation of vortex dipoles and the spread of the vortex
cluster should be visible in trapped atomic Bose-Einstein condensates
at sufficiently low temperatures. The presence of the thermal cloud near the
edge of the condensate prevents evaporating dipoles from
being re-injected into the vortex cluster and thus promotes a cleaner realization of vortex diffusion than at strictly zero temperature.

Our values of $\nu'/\kappa$ are slightly larger but in fair 
agreement with $\nu'/\kappa \approx 0.1$ reported by Tsubota {\it et al}
\cite{Tsubota2003},
and in the related (but not the same as spatial spreading) 
problem of the decay of superfluid turbulence 
\cite{Walmsley2014,Zmeev2015,Babuin2016}. The difference is probably 
due to the fact 
that our calculation is two-dimensional and lacks three-dimensional 
effects such as vortex reconnections and Kelvin waves. 
Further work will explore the problem in three dimensions 
using the GPE as well as the vortex filament model.

\begin{acknowledgements}
We are grateful with discussions with D. Proment.
NGP and CFB acknowledge the support of EPSRC grants EP/M005127/1 and
EP/R005192/1 respectively. This work used the facilities of N8 HPC 
provided and funded by the N8 consortium and EPSRC (grant EP/k000225/1).
\end{acknowledgements}

\clearpage

\end{document}